\documentstyle[12pt]{article}

\begin{document}

\baselineskip 24pt

\parindent 0pt

{\LARGE\bf Interfacial adsorption phenomena of the three-dimensional
 three-state Potts model}

\vspace{20pt}

{\Large Atsushi Yamagata}

Department of Physics, Tokyo Institute of Technology, Oh-okayama, Meguro-ku,
 Tokyo 152, Japan

\vspace{20pt}

Short title: Interfacial adsorption phenomena of the Potts model

Classification numbers: 64.60.Cn, 75.10.Hk

\vspace{20pt}

\parindent 1.5em

\parindent 0pt

{\bf Abstract.} We study the interfacial adsorption phenomena
 of the three-state ferromagnetic Potts model
 on the simple cubic lattice by the Monte Carlo method.
Finite-size scaling analyses of the net-adsorption yield
 the evidence of the phase transition being of first-order
 and $k_{\rm B} T_{\rm C} / J = 1.8166 (2)$.

\parindent 1.5em

\clearpage

\section{Introduction}

\label{sec : d3q31}

The Potts model is a multi-state statistical mechanical model
 for spin systems \cite{potts52,kms54,wu82}.
The Potts spin variable takes a value of 1, 2, $\ldots$ and $q$.
The Bragg-Williams approximation predicts that it exhibits
 a first-order phase transition for $q>2$ in all dimensions \cite{kms54}.

The properties of the two-dimensional models have been well understood.
According to the duality argument, the transition temperature is known
 for all $q$ exactly \cite{potts52,kms54}.
Baxter \cite{baxter73} proved the phase transition
 to be of first-order for $q>4$ and second-order for $q\le 4$.
There is the conjecture for the thermal and magnetic exponents $y_t$
 \cite{dennijs79} and $y_h$ \cite{pearson80,nrs80} for $q \leq 4$.

There is no exact result in three dimensions \cite{wu82}.
There is a problem whether the order of the transition
 of the three-dimensional three-state Potts model is of first-order or not.
It has been attacked by many authors and the result is that
 the system occurs a weakly first-order phase transition
 \cite{kj75,mbe79,bs79,herrmann79,kjm79,oi82,wv87,gkp89,fmou90,bonfim91,abv91}.
The transition is characterized by the small jump of the energy.

The specific heat shows the singularity of the delta function type
 as a result of the discontinuity of the energy at first-order phase
 transitions.
The behaviour can be seen only in systems with infinite lattice size.
In systems with finite lattice size the singularity of the energy is smoothed
 out and that of the specific heat is rounded off.
The finite-size scaling theory can predict behaviour of infinite systems
 through informations of finite systems \cite{fisher70,barber83,privman90}.

The Monte Carlo method is a useful tool to investigate phase transitions and
 critical phenomena \cite{binder79,bs87}.
Monte Carlo simulations are necessarily carried out on systems
 with finite lattice size.
With help of the finite-size scaling theory we can study thermodynamic
 behaviour of various physical systems.

The specific heat of the system with linear size $L$ has
 a maximum $C_{\rm max} (L)$ at a temperature $T_{\rm max}^C (L)$.
The finite-size effects are governed by the dimensions $d$
 at first-order phase transitions \cite{clb86,bkms91,lk91}:
\begin{displaymath}
C_{\rm max} (L) \sim L^d,
\end{displaymath}
\begin{displaymath}
T_{\rm max}^C (L) - T_{\rm C} \sim L^{-d}
\end{displaymath}
and by the critical exponents $\alpha$ and $\nu$ at second-order
 \cite{fisher70,barber83,privman90}:
\begin{displaymath}
C_{\rm max} (L) \sim L^{\alpha / \nu},
\end{displaymath}
\begin{displaymath}
T_{\rm max}^C (L) - T_{\rm C} \sim L^{- 1 / \nu}.
\end{displaymath}
The difference of the power has been used to determine the nature
 of transitions \cite{fmou90}.

As described above the five-state Potts model on the square lattice
 undergoes a first-order phase transition.
Peczak and Landau \cite{pl89} studied the system by the Monte Carlo method.
They found that it behaved as if the transition was of second-order,
 {\it i.e.} the physical quantities showed pseudocritical behaviour,
 {\it e.g.} $C_{\rm max} (L) \sim L^{1.09}$ rather than $L^2$.
Thus it is difficult to recognize the order of the transition.
Yamagata and Kasono \cite{yk92}, however, could found the finite-size effects
 of first-order type in the interfacial adsorption phenomena
 \cite{selke84,selke92} of the system by the Monte Carlo method.

The aim of this paper is to identify the transition of the three-state
 ferromagnetic Potts model on the simple cubic lattice and to determine
 the transition temperature by analysing the finite-size effects
 in the interfacial adsorption phenomena with the Monte Carlo method.
The phenomena in three dimensions are studied for the first time.
In the next section we describe the interfacial adsorption phanomena
 and then discuss the finite-size scaling theory for them
 in section \ref{sec : d3q33}.
We present the detail of our Monte Carlo simulations in section
 \ref{sec : d3q34}.
In section \ref{sec : d3q35} we analyse the Monte Carlo data.
A summary is given in section \ref{sec : d3q36}.

\section{Interfacial adsorption phenomena}

\label{sec : d3q32}

The interfacial adsorption phenomena \cite{selke84,selke92} have been
 observed on the multi-state models: the Potts models
 \cite{yk92,sp82,sh83,yamagata91}, the Blume-Capel model
 \cite{sy83,shk84} and the chiral clock model \cite{yd85,ps87}.
We shall explore them for the Potts models.

Let us consider the $q$-state ferromagnetic Potts model
 on the $L^{d-1} \times (L + 2)$ hypercubic lattice.
In the $d$th direction the fixed boundary conditions are taken and
 periodic boundary conditions are used in the remaining directions.
We fix the Potts spin variables in two states 1 and $m$, which takes the
 value of 1 or 2, at the opposite boundaries in the $d$th direciton,
 respectively.
The Hamiltonian is given by
\begin{equation}
{\cal H}_{(1|m)}
=-J\sum_{\langle i, j \rangle} \delta (\sigma_i, \sigma_j)
 -J{\sum_i}' \delta (\sigma_i, 1)
 -J{\sum_i}'' \delta (\sigma_i, m)
\label{eqn : hd3q3}
\end{equation}
where $\delta$ is the Kronecker's delta function;
$J(>0)$ is the strength of interactions;
the summation for $\langle i, j \rangle$ is over all nearest neighbour pairs
 on the lattice;
$\sum_i'$ and $\sum_i''$ denote the summations over lattice sites
 which adjoin to the opposite boundaries in the $d$th direction, respectively.

In the system ${\cal H}_{(1|2)}$ at a very low temperature an interface
 can appear between two phases 1 and 2 (called the $(1|2)$-interface).
It has been found that the remaining states 3, 4, $\ldots$ and $q$
 (called the non-boundary states) were generated in the form of droplets
 at the interface near the transition temperature $T_{\rm C}$ \cite{sp82,sh83}.
The interfacial adsorption phenomena are described by the net-adsorption
 defined \cite{sp82,sh83} as
\begin{equation}
W_L
\equiv L^{1-d} \sum_i \sum_{n=3}^q
\left[
\left\langle \delta (\sigma_i, n) \right\rangle_{(1|2)}
-\left\langle \delta (\sigma_i, n) \right\rangle_{(1|1)}
\right]
\label{eqn : defwd3q3}
\end{equation}
where $\langle \cdots \rangle_{(1|2)}$ and $\langle \cdots \rangle_{(1|1)}$
 denote the thermal or Monte Carlo averages for the systems ${\cal H}_{(1|2)}$
 and ${\cal H}_{(1|1)}$, respectively;
the summation for $i$ is over all lattice sites except those
 where the fixed Potts spin variables are located.
One should note that the summation for $n$ is over the non-boundary states.
The results of Monte Carlo simulations show that $W_L (T)$ has a finite peak
 near $T_{\rm C}$ on the finite system \cite{sp82}.

\section{Finite-size scaling theory}

\label{sec : d3q33}

Yamagata and Kasono \cite{yk92} have discussed the finite-size effects
 \cite{sh83,yamagata91} of the net-adsorption.
Noting the relation between $W_L (T)$ and the interfacial free energy,
 they have shown
\begin{equation}
W_L (T)
\approx L {\mit \Omega} (t L^d)
\label{eqn : scaw1st}
\end{equation}
and
\begin{equation}
W_L (T)
\approx L^{1-d+y_h} {\mit \Omega} (t L^{y_t})
\label{eqn : scaw2nd}
\end{equation}
at first- and second-order phase transitions, respectively,
 where ${\mit \Omega}$ is a scaling function, $t = 1 - T / T_{\rm C}$ and
 $y_t$ and $y_h$ are the thermal and magnetic exponents,
 respectively \cite{wu82,dennijs79,pearson80,nrs80}.
We omit the correction terms for simplicity here.
In two dimensions the finite-size scaling (\ref{eqn : scaw1st}) and
 (\ref{eqn : scaw2nd}) of $W_L (T)$ have been confirmed
 by the Monte Carlo method \cite{yk92,sh83,yamagata91}.
One should note in (\ref{eqn : scaw1st}) that $W_L (T_{\rm C})$
 is proportional to $L$ at first-order phase transitions regardless of the
 dimensions $d$.

Yamagata and Kasono \cite{yk92} have studied the finite-size effects
 of $W_L (T_{\rm C})$ to identify the order of transitions.
They have found by the Monte Carlo method that $W_L (T_{\rm C})$ was
 proportional to $L$ on the two-dimensional five-state Potts model.
It is clear evidence that the transition of the system is of first-order.
As described in section \ref{sec : d3q31}, the finite-size effect
 of the specific heat of the system could not be seen
 to be of first-order type.
It is important to choose physical quantities to be analysed.

We shall study the finite-size effects of the net-adsorption
 of the three-dimensional three-state Potts model on the same lines.
We, however, does not know the value of the transition temperature
 of the system exactly \cite{wu82}.
We note that $W_L (T)$ has a maximum $W_{\rm max} (L)$
 at a temperature $T_{\rm max}^W (L)$ on the system with linear size $L$.
Thus the scaling function ${\mit \Omega} (x)$ does so.
Let $x_{\rm max}$ be a location of the maximum of ${\mit \Omega} (x)$
 \cite{fl91}.
$x_{\rm max}$ satisfies a equation
\begin{displaymath}
\frac{{\rm d} {\mit \Omega}}{{\rm d} x} (x_{\rm max}) = 0.
\end{displaymath}

We consider the case of first-order phase transitions (\ref{eqn : scaw1st})
 hereafter.
Since it is clear that
\begin{displaymath}
x_{\rm max}
= \frac{T_{\rm C} - T_{\rm max}^W (L)}{T_{\rm C}} L^d,
\end{displaymath}
we get a relation
\begin{equation}
T_{\rm max}^W (L)
= T_{\rm C} - T_{\rm C} x_{\rm max} L^{-d}.
\label{eqn : tcwmaxd3q3}
\end{equation}
Since $x_{\rm max}$ is independent of $T$ and $L$, $W_{\rm max} (L)$ grows
 with $L$:
\begin{equation}
W_{\rm max} (L)
\equiv W_L (T_{\rm max}^W (L))
\sim L {\mit \Omega} (x_{\max}).
\label{eqn : wmaxd3q3}
\end{equation}
We may expect to estimate the transition temperature and to identify the
 nature of the transition by analysing the finite-size effects of the
 net-adsorption obtained by the Monte Carlo simulations.

\section{Monte Carlo simulations}

\label{sec : d3q34}

We use the Metropolis algorithm \cite{binder79,bs87} to simulate
 the systems ${\cal H}_{(1|2)}$ and ${\cal H}_{(1|1)}$ given
 by (\ref{eqn : hd3q3}) with $q=3$ on $L^2 \times (L+2)$ simple cubic lattice
 ($L$ = 10, 12, 14, 16, 18, 20, 24 and 30).
The net-adsorption is calculated with (\ref{eqn : defwd3q3}) of $d = q = 3$.
We start each simulation from a high temperature with a random
 configuration and then gradually cool the system.
The pseudorandom numbers are generated by the Tausworthe method
 \cite{tsuda88}.
Measurements at a temperature are over $10^5$ Monte Carlo steps
 per spin (MCS/spin) after discarding $10^4$ MCS/spin to attain equilibrium.
Near $T_{\rm max}^W (L)$ we observe the physical quantities
 over $10^6$ MCS/spin after $5 \times 10^4$ MCS/spin
 for the systems with $L$ = 14, 16, 18, 20, 24 and 30.
We have checked that simulations from the ground state configuration
 and a random one gave consistent results and there was no hysteresis.
We use the coarse-graining scheme to calculate the statistical errors
 \cite{landau76}.
Each run is devided into ten blocks and the standard deviations are obtained
 from the ten subaverages.

As described in the previous section, we want to estimate the maximum
 of the net-adsorption and its position for each lattice size.
Since it is difficult to get them from raw Monte Carlo data,
 we decide to adopt a procedure, the $B$-spline smoothing \cite{tsuda88}.
We fit our Monte Carlo data by the fourth-order $B$-spline.

\section{Monte Carlo results}

\label{sec : d3q35}

Hereafter, for brevity, the physical quantities are presented
 in units $k_{\rm B} = 1 = J$.

Figure 1 shows the temperature dependence of the net-adsorption $W_L (T)$
 defined by (\ref{eqn : defwd3q3}) with $d=q=3$ for various lattice sizes.
The net-adsorption has a finite peak.
We plot $W_{\rm max} (L)$ versus $L$ in figure 2.
It is clear that $W_{\rm max} (L)$ is proportional to $L$.
It is consistent with the prediction (\ref{eqn : wmaxd3q3})
 by the finite-size scaling theory.
By using the linear-regression from the data $W_{\rm max} (L)$
 with $L$ = 18, 20, 24 and 30 we obtain a relation:
 $W_{\rm max} (L) = 0.092(4) + 0.0382(2) L$.
In figure 3, $T_{\rm max}^W (L)$ is shown for the function of $L^{-3}$.
For systems with large lattice size it agrees well
 with (\ref{eqn : tcwmaxd3q3}).
We estimate $T_{\rm C}$ to be $1.8166 \pm 0.0002$
 by using the linear-regression from the data $T_{\rm max}^W (L)$
 with $L$ = 18, 20, 24 and 30; $T_{\rm max}^W (L) = 1.8166(2) -41(2) L^{-3}$.
The value should be compared with the previous results
 (see the table \ref{tbl : tcd3q3}).

\section{Summary}

\label{sec : d3q36}

We studied the interfacial adsorption phenomena of the three-state
 ferromagnetic Potts model on the simple cubic lattice.
The net-adsorption was calculated by the Monte Carlo method.
The finite-size effects were consistent with the predictions
 (\ref{eqn : tcwmaxd3q3}) and (\ref{eqn : wmaxd3q3})
 of the finite-size scaling theory for systems with $L \ge 18$.
It is the clear evidence of the phase transition being of first-order.
The transition temperature was estimated to be 1.8166(2).
It is consistent with the recent results \cite{fmou90,abv91}

The net-adsorption attains to the asymptotic region for $L \ge 18$.
The specific heat, on the other hand, shows the asymptotic behaviour
 for $L > 30$ \cite{fmou90}.
We succeeded in identifying the phase transition and getting
 the transition temperature from the data with $L \le 30$
 since we investigate the net-adsorption.
Generally we can get good statistics for small systems
 on the Monte Carlo simulations and it takes the long computer time
 to simulate large systems.
Thus we need to select the observational quantity carefully.

\section*{Acknowledgments}

The author would like to thank K. Kasono for the useful discussions
 and critical reading of the manuscript.
The simulations were carried out on the HITAC M-682 computer
 under the Institute of Statistical Mathematics Cooperative Research Program
 (92-ISM $\cdot$ CRP-45) and on the HITAC S-820/80 computer
 at the Computer center of Hokkaido University.
This study was supported by the Grant-in-Aid for Scientific Research
 from the Ministry of Education, Science and Culture, Japan.

\clearpage


\clearpage

\clearpage

\section*{Figure captions}

{\parindent 0pt

Figure 1.
The temperature dependence of the net-adsorption of the three-state Potts
 model on the simple cubic lattice
 with $L$ = 10, 12, 14, 16, 18, 20, 24 and 30.
The full curves are obtained by the $B$-spline smoothing.
As $L$ increases, the shape of the curve becomes sharper.

\vskip 24pt

Figure 2.
The size dependence of the maximum of the net-adsorption
 of the three-state Potts model on the simple cubic lattice.
The solid line shows $0.092 + 0.0382 L$.
Errors are less than symbol size.

\vskip 24pt

Figure 3.
The size dependence of the location of the net-adsorption maximum
 of the three-state Potts model on the simple cubic lattice.
The solid line shows $1.8166 - 41 L^{-3}$.
}

\clearpage

\begin{table}

\caption{The transition temperature of the three-state Potts model
 on the simple cubic lattice in units $k_{\rm B} = 1 = J$.}

\begin{center}

\begin{tabular}{ll}
\hline
\multicolumn{1}{c}{$T_{\rm C}$} & \multicolumn{1}{c}{Author(s)} \\
\hline
1.8166(2) & This work \\
1.816454(32) & Alves {\it et.al.} 1991 \\
1.8164(1) & Fukugita {\it et.al.} 1990 \\
1.8161(1) & Gavai {\it et.al.} 1989 \\
1.81624(6) & Wilson and Vause 1987 \\
1.81 & Ono and Ito 1982 \\
1.8169(6) & Knak Jensen and Mouritsen 1979 \\
1.818(3) & Herrmann 1979 \\
1.818 & Bl\"ote and Swendsen 1979 \\
1.827(1) &Miyashita {\it et.al.} 1979 \\
1.787(5) & Kim and Joseph 1975 \\
\hline
\end{tabular}

\end{center}

\label{tbl : tcd3q3}

\end{table}

\end{document}